\documentclass[aps,prl,floatfix,twocolumn]{revtex4}
\usepackage{amsmath}
\usepackage{amsfonts}
\usepackage{graphicx}
\usepackage{color}

\begin{document}

\title{Experimental demonstration of  Shor's algorithm with quantum entanglement \vspace{-2mm}} 
\author{B. P. Lanyon, T. J. Weinhold, N. K. Langford, M. Barbieri, D. F. V. James$^{*}$, A. Gilchrist, and A. G. White}
\affiliation{Centre for Quantum Computer Technology\hspace{1 mm}Department of Physics\hspace{1 mm}University of Queensland, Brisbane QLD 4072, Australia\\
$^{*}$Department of Physics\hspace{1mm}Center for Quantum Information and Control
\hspace{1 mm}University of Toronto, Toronto ON M5S1A7, Canada
\vspace{-2 mm}
}
\begin{abstract} \noindent Shor's powerful quantum algorithm for factoring represents a major challenge in quantum computation and its full realization will have a large impact on modern cryptography. Here we implement a compiled version of Shor's algorithm in a photonic system using single photons and employing the non-linearity induced by measurement. For the first time we demonstrate the core processes, coherent control, and resultant entangled states that are required in a full-scale implementation of Shor's algorithm. Demonstration of these processes is a necessary step on the path towards a full implementation of Shor's algorithm and  scalable quantum computing.  Our results highlight that the performance of a quantum algorithm is not the same as performance of the underlying quantum circuit, and stress the importance of developing techniques for characterising quantum algorithms.
\end{abstract}
\maketitle

\noindent As computing technology rapidly approaches the nano-scale, fundamental quantum effects threaten to introduce an inherent and unavoidable source of noise. An alternative approach embraces quantum effects for computation. Algorithms based on quantum mechanics allow tasks impossible with current computers, notably an exponential speed-up in solving problems such as the factoring problem \cite{Shor}. Many current cryptographic protocols rely on the computational difficulty of finding the prime factors of a large number: a small increase in the size of the number leads to an exponential increase in computational resources. Shor's quantum algorithm for factoring composite numbers faces no such limitation, and its realization represents a major challenge in quantum computation. 

To date, there have been demonstrations of entangling quantum-logic gates in a range of physical architectures, ranging from trapped ions \cite{Ions2, Ions3}, to superconducting circuits \cite{SC}, to single photons \cite{KLMexp,clusterexp,JWPan,Tangle,Weinfurter,Takeuchi,AccurateTomo,Nature07}. Photon polarisation experiences essentially zero decoherence in free space; uniquely, photonic gates have been fully characterised \cite{AccurateTomo}, produced the highest entanglement \cite{Tangle}, and are the fastest of any architecture \cite{Nature07}. The combination of long decoherence time and fast gate speeds make photonic architectures a promising approach for quantum computation, where large numbers of gates will need to be executed within the coherence lifetime of the qubits.

Shor's algorithm can factor a $k$-bit number using $72k^3$ elementary quantum gates, e.g.\ factoring the smallest meaningful number, $15$, requires $4608$ gates operating on $21$ qubits \cite{Preskill}. This is well beyond the reach of current technology. Recognizing this, Ref.~\cite{Preskill} introduced a compiling technique which exploits properties of the number to be factored, allowing exploration of Shor's algorithm with a vastly reduced number of resources. Although the implementation of these compiled algorithms do not directly imply scalability, they do allow the characterisation of core processes required in a full-scale implementation of Shor's algorithm. Demonstration of these processes is a necessary step on the path towards scalable quantum computing. These processes include the ability to generate entanglement between qubits by coherent application of a series of quantum gates: this represents a significant challenge with current technology. In the only demonstration to date, a compiled set of gate operations were implemented in a liquid NMR architecture \cite{ShorNMR}. However, since the qubits are at all times in a highly mixed state \cite{Caves}, and the dynamics can be fully modelled classically \cite{Menicucci},  neither the entanglement nor the coherent control at the core of Shor's algorithm can be implemented or verified.

Here we implement a compiled version of Shor's algorithm, using photonic quantum-logic gates to realise the necessary processes, and verify the resulting entanglement via quantum state and process tomography \cite{James,QPT}. We use a linear-optical architecture where the required nonlinearity is  induced by measurement; current experiments are not scalable, but there are clear paths to a fully scalable quantum architecture \cite{KLMth,clusterth}. Our gates do not require pre-existing entanglement and we encode our qubits into the polarisation of up to four photons. Our results highlight that the performance of a quantum algorithm is not the same as performance of the underlying quantum circuit, and stress the importance of developing techniques for characterising quantum algorithms.

Only one step of Shor's algorithm to find the factors of a number $N$ requires a quantum routine. Given a randomly chosen co-prime $C$ (where 1$<$$C$$<$$N$ and the greatest common divisor of $C$ and $N$ is 1), a quantum routine finds the \emph{order} of $C$ modulo $N$, defined to be the minimum integer $r$ that satisfies the function $C^{r} \mathrm{mod} N{=}1$. It is straightforward to find the factors from the order.  Consider $N{=}15$: if we choose $C$=2, the quantum routine finds $r{=}4$, and the prime factors are given by the non-trivial greatest common divisor of $C^{r/2}{\pm}1$ and $N$, i.e. 3 and 5; similarly if we choose the next possible co-prime, $C$=4, we find the order $r$=2, yielding the same factors.
\begin{figure}[!b]
\vspace{-6.5 mm}
\includegraphics[width=0.75 \columnwidth]{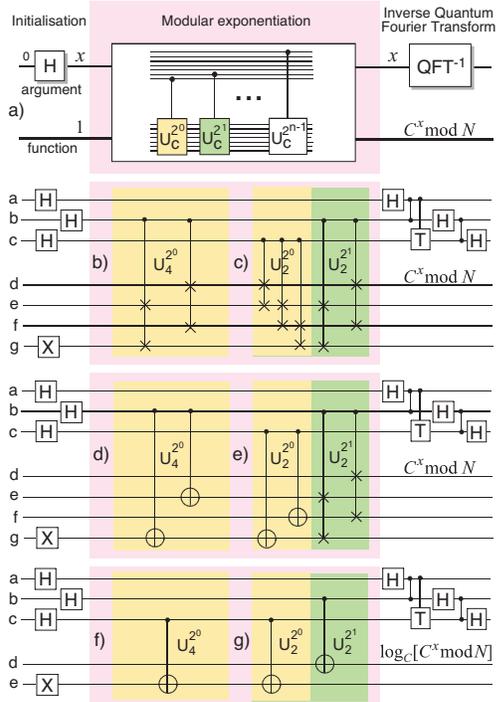}
\vspace{-3.5 mm}
\caption{a) Conceptual circuit for the order-finding routine of Shor's algorithm for number $N$ and co-prime $C$ \cite{Preskill}. The argument and function registers are bundles of $n$ and $m$ qubits; the nested order-finding structure uses $U | y \rangle$=$| C y\, \mathrm{mod} N \rangle$, where the initial function-register state is $| y \rangle{=}1$. The algorithm is completed by logical measurement of the argument-register, and reversing the order of the argument qubits. b),c) Implementation of a) for $N{=}15$ and $C{=}4, 2$, respectively; the unitaries are decomposed into controlled-\textsc{swap} gates (\textsc{cswap}), marked as $\textsc{x}$; controlled-phase gates are marked by dots; \textsc{h} and \textsc{t} represent Hadamard and $\pi/8$ gates. Many gates are redundant, e.g. the second gate in b), the first and second gates in c). d),e) Partially-compiled circuits of b),c), replacing \textsc{cswap} by controlled-\textsc{not} gates. n.b. e) is equivalent to the $N{=}15$ $C{=}7$ circuit in Ref.\cite{ShorNMR}. f),g) Fully-compiled circuits of d),e), by evaluating $\log_{C} [C^{x}\mathrm{mod}N]$ in the function-register.}
\label{fig:concept}
\end{figure}

Fig.~\ref{fig:concept}a) shows a conceptual circuit of the quantum order-finding routine. It consists of three distinct steps: i) \emph{register initialisation}, $| 0 \rangle^{\otimes n} | 0 \rangle^{\otimes m}{\rightarrow}(| 0 \rangle{+}| 1 \rangle)^{\otimes n}$ $| 0 \rangle^{\otimes m{-}1}| 1 \rangle {=}\sum_{x{=}0}^{2^{n}{-}1} | x \rangle| 0 \rangle^{\otimes m{-}1}| 1 \rangle$, where the argument-register is prepared in an equal coherent superposition of all possible arguments (normalisation omitted by convention); ii) \emph{modular exponentiation}, which by controlled application of the order-finding function produces the entangled state $\sum_{x=0}^{2^{n}-1} | x \rangle | C^{x} \mathrm{mod} N \rangle$; iii) the \emph{inverse Quantum Fourier Transform} (QFT) followed by measurement of the argument-register in the logical basis, which with high probability extracts the order $r$ after further classical processing. If the routine is standalone, the inverse QFT can be performed using an approach based on local measurement and feedforward \cite{GN}. Note that the inverse QFT in \cite{ShorNMR} was unnecessary: it is straightforward to show this is true for any order-$2^{l}$ circuit  \cite{epaps}. 

\begin{figure}[!b]
\vspace{-7 mm}
\includegraphics[width=0.8\columnwidth]{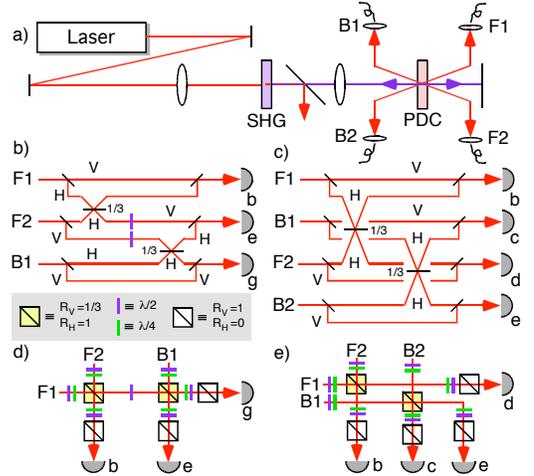}
\vspace{-3 mm}
\caption{Experimental schematic. a) Forward and backward photons pairs are produced via parametric downconversion (PDC) of a frequency-doubled mode-locked Ti:Sapphire laser (820 nm$\rightarrow$410 nm, $\Delta \tau{=}80$ fs at 82 MHz repetition rate) through a Type-I 2 mm Bismuth Borate (BiB$_{3}$O$_{6}$) crystal. Photons are input to the circuits via blocked interference filters (820$\pm$3 nm) and single-mode optical fibres, and detected using single photon counting modules, (PerkinElmer AQR-14FC). Coincidences are measured using a quad-logic card driven by a four-channel constant fraction discriminator. With 500 mW at 410 nm this yielded 60 kHz and 25 kHz twofold coincidence rates for direct detection, which differed due to mismatched pump focus sizes; the measured fourfold coincidence rate was 35 Hz. b),c) Linear optical circuits for order-2 and order-4 finding algorithms, with inputs from a) labelled; the letters on the detectors refer to the Fig.~1 qubits. d),e) Physical optical circuits for b),c), replacing the classical interferometers with partially-polarising beamsplitters.}
\label{fig:circuit}
\end{figure}

Modular exponentiation is the most computationally-intensive part of the algorithm \cite{Preskill}. It can be realised by a cascade of controlled unitary operations, $U$, as shown in the nested inset of Fig.~1a). It is clear that the registers become highly entangled with each other: since $U$ is a function of $C$ and $N$, the entangling operation is unique to each problem. Here we choose to factor 15 with the first two co-primes, $C{=}2$ and $C{=}4$. In these cases entire sets of gates are redundant: specifically, $U^{2^{n}}{=}I$ when $n{>}0$ for $C{=}4$, and $U^{2^{n}}{=}I$ when $n{>}1$ for $C{=}2$. Figs~1b),c) show the remaining gates for $C{=}4$ and $C{=}2$, respectively, after decomposition of the unitaries into controlled-\textsc{swap} gates---this level of compiling is equivalent to that introduced in Ref.~\cite{ShorNMR}. Further compilation can always be made since the initial state of the function-register is fixed, allowing the \textsc{cswap} gates to be replaced by controlled-\textsc{not} (\textsc{cnot}) gates as shown in Figs~1d),e) \cite{Note2}. 

We implemented the order-2-finding circuit, Fig.~1d). The qubits are realised with simultaneous forward and backward production of photon pairs from parametric downconversion, Fig.~\ref{fig:circuit}a): the logical states are encoded into the vertical and horizontal polarisations. This circuit required implementing a recently-proposed three-qubit quantum-logic gate, Fig.~\ref{fig:circuit}b), which realises 
a cascade of $n$ controlled-\textsc{z} gates with exponentially greater success than chaining $n$ individual gates \cite{Ralph}. The controlled-\textsc{not} gates are realised by combining Hadamards and controlled-\textsc{z} gates based on partially-polarising beamsplitters. The gates are nondeterministic, with one third success probability when fully prebiased \cite{Tangle,Weinfurter,Takeuchi}. A run of each routine is flagged by a fourfold event, where a single photon arrives at each output.  Dependent photons from the forward pass interfere non-classically at the first partial polariser, Fig.~\ref{fig:circuit}d), one photon then interferes with an independent photon from the backward pass at the second partial polariser. We measured relative nonclassical visibilities, $V_{r}{\equiv}V_{\mathrm{meas}}/V_{\mathrm{ideal}}$, of $98{\pm}2\%$ and $85{\pm}6\%$.

Directly encoding the order-4 finding circuit, Fig.~1e), requires six photons and at least one three-qubit and five two-qubit gates. This is currently infeasible:  the best six-photon rate to date \cite{JWPan} is 30 mHz, which would be reduced by six orders of magnitude using non-deterministic gates. To explore an order-4 routine, and the different processes therein, further compilation is necessary. In particular, we can compile circuits 1d),e) by evaluating $\log_C\left[C^x \mathrm{mod} N\right]$ in the function-register in place of $C^x \mathrm{mod} N$. This requires $\log_{2}[\log_{C}[N]]$ function qubits, as opposed to $\log_{2}[N]$, i.e. for N=15, C=2, the function-register reduces from 4 to 2 qubits. Note that this full compilation maintains all the features of  the algorithm as originally proposed in Ref.~\cite{Preskill}.
Thus the order-4 circuit, Fig.~\ref{fig:concept}e), reduces to a pair of \textsc{cnot}s, allowing us to implement the circuit in Fig.~\ref{fig:concept}g). We use a pair of compact optical gates \cite{Tangle,Weinfurter,Takeuchi}, Fig~\ref{fig:circuit}c),e), each operating on a dependent pair of photons, resulting in measured visibilities for both of $V_{r}{=}$98$\pm$2\%. 
\begin{figure}[!b]
\vspace{-7 mm}
\includegraphics[width=0.7 \columnwidth]{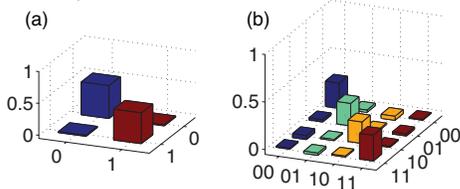}
\vspace{-5.5 mm}
\caption{Algorithm outputs given by measured argument-register density matrices. The diagonal elements are the logical output probabilities.
a) Order-2 algorithm. The fidelity with the ideal state is $F{=} 99.9{\pm}0.3\%$,
the linear entropy is $S_{L}{=}100{\pm}1\%$ \cite{josaBFeb}. Combined with the redundant qubit the logical probabilities are $\{P_{00},P_{10}\}{=}\{52,48\}{\pm}3\%$.
b) Order-4 algorithm, $F{=}98.5{\pm}0.6\%$, and $S_{L}{=}98.1{\pm}0.8\%$. The logical probabilities are  $\{P_{000},P_{010},P_{100},P_{110}\}{=}\{27,23,24,27\}{\pm}2\%$. Real parts shown, imaginary parts are less than $0.6\%$.}
\label{fig:algo}
\end{figure}
\begin{figure}[!b]
\vspace{-8 mm}
\includegraphics[width=0.78 \columnwidth]{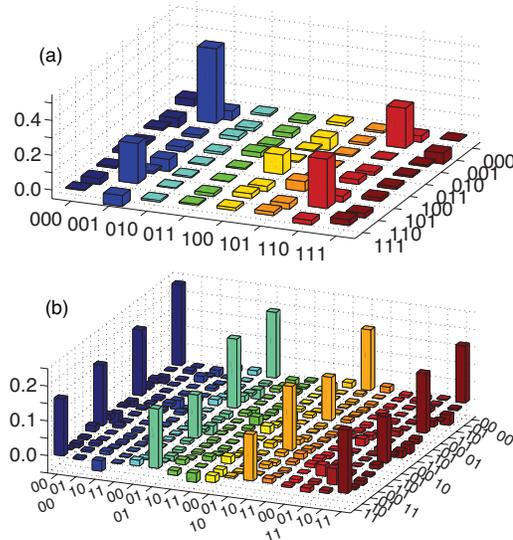}
\vspace{-5.5 mm}
\caption{Measured density matrices of the state of both registers after modular exponentiation. 
a) Order-2 circuit. Ideal state is locally equivalent to a GHZ state: we find $F_{\mathrm{GHZ}}{=}59{\pm}$4\%. The state is partially-mixed, $S_{L}$=62\%$\pm$4\%, and entangled, violating the optimal GHZ entanglement witness $W_{\mathrm{GHZ}}{=}1/2{-}F_{\mathrm{GHZ}}{=}{-9}{\pm}4$\% \cite{GHZWitness}.
b) Order-4 circuit. Measured fidelity with the ideal state, a tensor product of two Bell-states, is $F{=}68{\pm}3\%$. The state is partially-mixed, $S_{L}{=}52{\pm}4\%$, and entangled, with tangles of the component Bell-States of $41{\pm}5\%$ and $33{\pm}5\%$. Real parts shown, imaginary parts are respectively less than $7\%$ and $4\%$.}
\label{fig:modexp}
\end{figure}

Fig.~\ref{fig:algo} shows the measured density matrices of the argument-register output for both algorithms, sans the redundant top-rail qubit \cite{Note3}. Ideally these are maximally-mixed states \cite{epaps}: in all cases we measure near-unity fidelities \cite{FidNote,josaBFeb}. The output of the routines are the logical state probabilities, i.e. the diagonal elements of the matrices. Combining these with the known state of the redundant qubit, and reversing the argument qubits as required, gives the binary outputs of the algorithm which after classical processing yields the prime factors of $N$. In the order-2 circuits the binary outputs of the algorithm are 00 or 10:  the former represents the expected failure mode of this circuit, the latter a successful determination of $r{=}2$; failure and success should have equal probabilities, we measure them to be 50\% to within error. Thus half the time the algorithm yields $r{=}2$, which gives the factors, 3 and 5. In the order-4 circuit the binary outputs are 000, 010, 100 and 110: the second and fourth terms yield the order-4 result, the first is a failure mode and the third yields trivial factors. We measure output probabilities of 25\% to within error, as expected. After classical processing half the time the algorithm finds $r{=}4$, again yielding the factors 3 and 5.

These results show that we have near-ideal algorithm performance, far better than we have any right to expect given the known errors inherent in the logic gates \cite{Tangle,Till}. This highlights that the \emph{algorithm} performance is not always an accurate indicator of \emph{circuit} performance since the algorithm produces mixed states. In the absence of the gates the argument-register qubits would remain pure; as they are mixed they have become entangled to \emph{something} outside the argument-register. From algorithm performance we cannot distinguish between desired mixture arising from entanglement with the function-register, and undesired mixture due to environmental decoherence. Circuit performance is crucial if it is to be incorporated as a sub-routine in a larger algorithm, Fig.~1a), e), and g). The \emph{joint} state of both registers after modular exponentiation indicates circuit performance; we find entangled states that partially overlap with the expected states, Fig.~\ref{fig:modexp}, indicating some environmental decoherence. The fidelity of the four-qubit state with the ideal, Fig.~\ref{fig:modexp}b), is higher than that of the three-qubit state, Fig.~\ref{fig:modexp}a), chiefly because the latter requires nonclassical interference of photons produced by independent sources, which suffer higher distinguishability, lowering gate performance \cite{Till,rarity,IP2}.

Process tomography fully characterises circuit performance, yielding the $\chi$-matrix, a table of process measurement outcomes and the coherences between them. Measured and ideal $\chi$-matrices can be quantitatively compared using the fidelity \cite{AccurateTomo,josaBFeb}; we measured process fidelities of $F_{p}{=}85\%, 89\%$ for the two-qubit gates of the order-4 circuit. It is the easier of the two algorithms to characterise since it consists of two gates acting on independent qubit pairs. Consequently, by assuming that only these gates induce error, the order-4 circuit process fidelity is simply the product of the individual gate fidelities \cite{Raginsky}, $F_{p}^{bcde}{=}F_{p}^{bd}F_{p}^{ce}{=}80\%$. 
Clearly this is significantly less than the \emph{algorithm} success rate of 99.7\%. 
The order-2 circuit is harder to characterise, requiring at least 4096 measurements, infeasible with our count rates.  Decomposing the three-qubit gate into a pair of two-qubit gates yields process fidelities $F_{p}{=}78\%, 90\%$ (again reflecting differing interferences of independent and dependent photons).
There is no simple relation between individual \textsc{cz} gate performances, and that of the three-qubit gate. However, a bound can be obtained by chaining the gate errors, $F_{p} \ge$20\% \cite{05gilchrist}. This is not a useful bound, c.f. the fidelity between an ideal \textsc{cz} and doing nothing at all of $F_{p}{=}25\%$! (The bound only becomes practical as $F_{p}{\rightarrow}1$). For larger circuits, full tomographic characterisation becomes exponentially impractical. The order-finding routine registers contain $k{=}n{+}m$ qubits: state and process tomography of a $k$-qubit system require at least $2^{2k}$ and $2^{4k}$ measurements, respectively.

\begin{figure}[!t]
\includegraphics[width=0.75 \columnwidth]{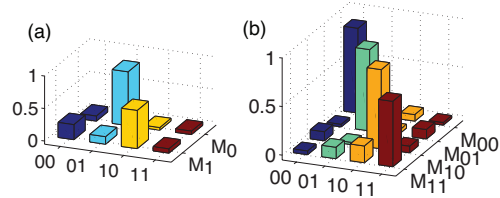}
\vspace{-5 mm}
\caption{Measured function-register probabilities after modular exponentiation, conditioned on logical measurement of the argument-register $M_{x}$. There is a high correlation between the registers: a) Order-2 circuit, $\{P_{01},P_{10}\}{=}\{83{\pm}4\%,59{\pm}5\%\}$; b) Order-4 circuit, $\{P_{00},P_{01},P_{10},P_{11}\}{=}\{87{\pm}3\%,84{\pm}4\%,82{\pm}5\%,67{\pm}6\%\}$. \vspace{-7 mm}}
\label{fig:func reg}
\end{figure}

An alternative is to gauge circuit performance via the logical correlations \emph{between} the registers. Modular exponentiation produces the entangled state $\sum_{x=0}^{2^{n}{-}1} | x \rangle | y \rangle$, where $y$ is respectively $C^{x} \mathrm{mod} N$ and $\log [C^{x} \mathrm{mod} N]$ for partial and full compilation. For a correctly functioning circuit, measuring the argument in the state $x$ projects the function into $y$---requiring at most $2^{k}$ measurements to check. The results in Fig.~\ref{fig:func reg} show there is a clear correlation between the argument and function registers, 59 to 83\% and 67 to 87\% for the order-2 and order-4 circuits, respectively. Again, these indicative values of circuit operation are significantly less than the algorithm success rates. 

We have experimentally implemented every stage of a small-scale quantum algorithm. Our experiments demonstrate the feasibility of executing complex, multiple-gate quantum circuits involving coherent multi-qubit superpositions of data registers. We present two different implementations of the order-finding routine at the heart of Shor's algorithm, characterising the algorithmic and circuit performances. 
Order-finding routines are a specific case of phase-estimation routines, which in turn underpin a wide variety of quantum algorithms, such as those in quantum chemistry \cite{QChem}.  Besides providing a proof of the use of quantum entanglement for arithmetic calculations, this work points to a number of interesting avenues for future research---in particular, the advantages of tailoring algorithm design to specific physical architectures, and the urgent need for efficient diagnostic methods of large quantum information circuits.

We thank M.\ P.\ de Almeida and E.\ DeBenedictis for discussions, and the Australian Research Council, US Disruptive Technologies Office (Contract W911NF-05-0397), Canadian National Science and Engineering Research Council and DEST Endeavour Europe Awards, for support. 

\newpage

\newpage

\noindent 
\textbf{Additional Online Material}. For all the circuits Fig.~1
b)-g), the consecutive Hadamards in the top qubit of the argument-register cancel each other out (since $\textsc{h}^{2}{=}\textsc{i}$): consequently both this qubit, and the gate(s) controlled by it, are redundant and need not be implemented experimentally. The remaining argument-register qubits are maximally-entangled to the function-register. Since the function-register output is not measured, these argument qubits are maximally-mixed, and the subsequent gates in the inverse QFT are therefore also redundant. Thus the inverse QFT in Ref.~[14] 
was unnecessary: indeed, it is straightforward to show this is true for any order-$2^{l}$ circuit. 
After modular exponentiation, the circuit state is $\sum_{x=0}^{2^n-1}|x\rangle|C^x\mathrm{mod}N\rangle$: for any two values $x$ and $y$ that differ by an integer, $k$ number of orders, i.e. $y{-}x{=}k \, 2^{l}$, $C^y \mathrm{mod} N {=} C^{x} \mathrm{mod} N$, and the state after modular exponentiation becomes  $\sum_{k=0}^{2^{n-l}-1}\sum_{a=0}^{2^{l}-1} | k 2^{l}{+}a \rangle | C^{a} \mathrm{mod}N \rangle$. Note that the first $n{-}l$ qubits of the argument-register (top to bottom) encode the number $k$, the remaining $l$ qubits encode $2^l$ distinct values of $a$: we divide the argument-register accordingly, $\sum_{k, a} | k \rangle | a \rangle | C^a\rangle$. The $|k\rangle$ qubits do not become entangled to the function-register whereas the $|a\rangle$ qubits are maximally-entangled to it---consequently after tracing out the function-register, the $|a\rangle$ qubits are in a maximally-mixed state and any further gates acting on them are redundant.
Application of Hadamard gates in the inverse QFT reset the $|k\rangle$ qubits to 0, inhibiting any gates controlled by them, The final step of the inverse QFT is to swap the first and last qubits of the argument  register which can be done after measurement. Thus the inverse QFT can be omitted in all cases $r{=}2^l$.


\vspace{-7 mm}
\begin{thebibliography}{99}

\bibitem{Shor} P. Shor, \emph{Proc.35th Ann. Symp. Found. Comp. Sci.}, 124 (IEEE Comp.Soc.Press, Los Alamitos, California, 1994).

\bibitem{Ions2} F. Schmidt-Kaler \emph{et al.}, \emph{Nature} {\bf 422}, 408 (2003).

\bibitem{Ions3} D. Leibfried \emph{et al.}, \emph{Nature} {\bf 422}, 412 (2003).

\bibitem{SC} M. Steffen \emph{et al.}, \emph{Science} {\bf 313}, 1423 (2006).

\bibitem{KLMexp} J.~L. O'Brien, \emph{et al.}, \emph{Nature} {\bf 426}, 264 (2003).

\bibitem{AccurateTomo} J. L. O'Brien \emph{et al.}, \emph{Phys. Rev. Lett.} {\bf 93}, 080502 (2004).

\bibitem{clusterexp} P. Walther \emph{et al.}, \emph{Nature} {\bf 434}, 169 (2005).

\bibitem{Tangle} N.~K. Langford, \emph{et al.}, \emph{Phys. Rev. Lett.} {\bf 95}, 210504 (2005).

\bibitem{Weinfurter} N. Kiesel {\emph{et al.}}, \emph{Phys. Rev. Lett.} {\bf 95}, 210505 (2005).

\bibitem{Takeuchi} R. Okamoto \emph{et al.}, \emph{Phys. Rev. Lett.} {\bf 95}, 210506 (2005).

\bibitem{Nature07} R. Prevedel \emph{et al.}, \emph{Nature} {\bf 445}, 65 (2007). 

\bibitem{JWPan} C.-Y. Lu \emph{et al.}, \emph{Nature Physics} {\bf 3}, 91-95 (2007).

\bibitem{Preskill} D. Beckman \emph{et al.}, \emph{Phys. Rev. A}. \textbf{54}, 1034 (1996).

\bibitem{ShorNMR} L.~M.~K. Vandersypen \emph{et al.}, \emph{Nature}  {\bf 414}, 883 (2001).

\bibitem{Caves} S. L. Braunstein \emph{et al.}, \emph{Phys. Rev. Lett.} \textbf{83}, 1054(1999).

\bibitem{Menicucci} N.C.Menicucci \emph{et al.}, \emph{Phys. Rev. Lett.} {\bf 88}, 167901 (2002).

\bibitem{James}  D.~F.~V. James \emph{et al.}, \emph{Phys. Rev. A}  {\bf 64}, 052312 (2001).

\bibitem{QPT} J. F. Poyatos, \emph{et al.}, \emph{Phys. Rev. Lett.} {\bf 78}, 390-393 (1997).

\bibitem{KLMth} E. Knill, \emph{et al.}, \emph{Nature} {\bf 409}, 46 (2001).

\bibitem{clusterth} M. A. Nielsen, \emph{Phys. Rev. Lett.}, {\bf 93}, 040503 (2004).

\bibitem{GN} R. B. Griffiths \emph{et al.}, \emph{Phys. Rev. Lett.} {\bf 76}, 3228 (1996).

\bibitem{epaps} See additional online material.

\bibitem{Note2} Fig.~1e) is equivalent to the order-4 $C{=}7$ circuit in Ref.~\cite{ShorNMR}: \textsc{cswap} is equivalent to a Toffoli and \textsc{cnot}s.

\bibitem{Ralph} T. C. Ralph, \emph{Phys. Rev. A} {\bf 70} 012312 (2004).

\bibitem{Note3} We use convex optimisation tomography \cite{QPTconvex} to estimate errors via Monte-Carlo simulation \cite{AccurateTomo}.  

\bibitem{QPTconvex} 
A. Doherty and A. Gilchrist, in preparation (2007).  

\bibitem{FidNote} Fidelity is $F(\rho,\sigma){\equiv}\mathrm{Tr}[\sqrt{\sqrt{\rho} \sigma \sqrt{\rho}}]^{2}$; linear entropy is $S_{L} {\equiv}$ $d (1{-}\mathrm{Tr}[\rho^2])/(d{-}1)$, where $d$ is the state dimension \cite{josaBFeb}.

\bibitem{josaBFeb} A. G. White \emph{et al.}, \emph{J. Opt. Soc. Am. B} {\bf 24}, 172 (2007).

\bibitem{Till} T. J. Weinhold \emph{et al.}, in preparation, (2007).

\bibitem{rarity} J. G. Rarity \emph{et al.}, \emph{J. Opt. B.} \textbf{7} S171 (2005).

\bibitem{IP2} R.~Kaltenbaek \emph{et al.}, \emph{Phys. Rev. Lett.} {\bf 96}, 240502 (2006).

\bibitem{Raginsky} M. Raginsky, \emph{Physics Letters} A {\bf 290}, 11-18 (2001).

\bibitem{05gilchrist} A. Gilchrist, \emph{et al.}, \emph{Phys. Rev. A} {\bf 71}, 062310 (2005).

\bibitem{QChem} A. Aspuru-Guzik, \emph{et al.}, \emph{Science} \textbf{309}, 1704-1707 (2005).

\bibitem{GHZWitness} M. Bourennane \emph{et al.}, \emph{Phys. Rev. Lett.} {\bf 92} 087902 (2004).

\end{thebibliography}
\end{document}